\documentclass[11pt]{article}
\usepackage{amsmath}
\usepackage{pstricks}

\def\half{\frac{1}{2}}

\newfont{\bbbold}{msbm10 scaled \magstep1}

\def\bbF{\mbox{\bbbold F}}

\def\bbR{\mbox{\bbbold R}}

\def\bbZ{\mbox{\bbbold Z}}
\def\cA{{\cal A}}

\def\cH{{\cal H}}

\def\cL{{\cal L}}

\def\cN{{\cal N}}
\def\cO{{\cal O}}

\def\cV{{\cal V}}

\newfont{\goth}{eufm10 scaled \magstep1}

\def\gg{\mbox{\goth g}}

\def\gl{\mbox{\goth l}}

\def\go{\mbox{\goth o}}
\def\gp{\mbox{\goth p}}

\def\gs{\mbox{\goth s}}

\def\a{\alpha}
\def\b{\beta}
\def\c{\gamma}\def\C{\Gamma}
\def\d{\delta}
\def\ve{\varepsilon}
\def\f{\phi}\def\F{\Phi}

\def\r{\rho}

\def\t{\tau}

\def\be{\begin{equation}}\def\ee{\end{equation}}
\def\bea{\begin{eqnarray}}\def\eea{\end{eqnarray}}
\def\barr{\begin{array}}\def\earr{\end{array}}

\def\o{\omega}\def\O{\Omega}

\def\xz{\times}

\def\nab{\nabla}

\let\la=\label

\def\nn{\nonumber}
\def\bd{\begin{document}}
\def\ed{\end{document}}
\def\ba{\begin{array}}
\def\ea{\end{array}}
\def\bea{\begin{eqnarray}}
\def\eea{\end{eqnarray}}
\def\ft#1#2{\tfrac{#1}{#2}}
\def\fft#1#2{\frac{#1}{#2}}
\def\sst#1{{\scriptscriptstyle #1}}
\def\oneone{\rlap 1\mkern4mu{\rm l}}
\def\bref{\blue{ref}}

\newcommand{\eq}[1]{(\ref{#1})}
\newcommand{\w}[1]{\\[0.#1cm]}
\def\eqs#1#2{(\ref{#1}-\ref{#2})}
\def\det{{\rm det\,}}
\def\tr{{\rm tr}}
\def\ad{{\rm ad}}

\newcommand{\hoch}[1]{$\, ^{#1}$}
\newcommand{\imperial}{\it\small Theoretical Physics Group, Imperial College London\\ Prince Consort Road, London SW7 2AZ, UK}
\newcommand{\kings}
{\it\small Department of Mathematics, King's College, University of London\\ Strand, London WC2R 2LS, UK}
\newcommand{\uu}
{\it\small Department of Theoretical Physics, Uppsala, Sweden}
\newcommand{\hip}
{\it\small HIP-Helsinki Institute of Physics, P.O. Box 64 FIN-00014
University of Helsinki, Suomi-Finland}
\newcommand{\stock}
{\it\small Department of Theoretical Physics, Stockholm, Sweden}
\newcommand{\golm}
{\it\small AEI, Max Planck Institut f\"ur Gravitationsphysik\\ Am M\"{u}hlenberg 1, D-14476 Potsdam, Germany}
\makeatletter
\renewcommand\theequation{\thesection.\arabic{equation}}
\@addtoreset{equation}{section} \makeatother

\newcommand{\sa}{/ \hspace{-1.2ex}}
\newcommand{\saa}{/ \hspace{-1.4ex}}
\newcommand{\saaa}{\, / \hspace{-1.6ex}}
\newcommand{\Scal}[1]{\Bigl ({#1} \Bigr )}
\newcommand{\scal}[1]{\bigl ({#1} \bigr )}

\newcommand{\CR}{\nonumber \\*}

\newcommand{\trace}{\hbox {tr}~}
\newcommand{\traceS}{\hbox {tr}_{\scriptscriptstyle \mathfrak{S}}~}

\DeclareMathAlphabet{\mathpzc}{OT1}{pzc}{m}{it}
\def\BRST{\,\mathpzc{s}\,}
\def\aBRST{{\scriptstyle (\mathpzc{s})}}
\def\q{{{\scriptscriptstyle (Q)}}}
\def\qs{{\scriptscriptstyle (Q\mathpzc{s})}}
\def\Qsla{{\mathcal{S}_{\q}}}
\def\Slav{{\mathcal{S}_\aBRST}}
\def\epsilonb{{\overline{\epsilon}}}
\def\bulletup{{\scriptstyle \bullet}}

\newcommand{\gra}[2]{{\scriptscriptstyle (#1 , #2 )}}
\newcommand{\ord}[1]{{\scriptscriptstyle (#1)}}

\def\cL{{\cal L}}
\def\cN{\mathcal{N}}
\def\cO{\mathcal{O}}

\def\ie{{\it i.e.}\ }
\def\eg{{\it e.g.}\ }

\newcommand{\sfrac}[2]{{\scriptstyle \frac{#1}{#2}}}
\newcommand{\stfrac}[2]{{\scriptscriptstyle \frac{#1}{#2}}}

 \def\balpha{{\overline{\alpha}}}
 \def\bbeta{{\overline{\beta}}}
 \def\bgamma{{\overline{\gamma}}}
 \def\bdelta{{\overline{\delta}}}
 \def\bepsilon{{\overline{\epsilon}}}
 \def\bvarepsilon{{\overline{\varepsilon}}}
 \def\bzeta{{\overline{\zeta}}}
 \def\bareta{{\overline{\eta}}}
 \def\btheta{{\overline{\theta}}}
 \def\bvartheta{{\overline{\vartheta}}}
 \def\biota{{\overline{\iota}}}
 \def\bkappa{{\overline{\kappa}}}
 \def\blambda{{\overline{\lambda}}}
 \def\bmu{{\overline{\mu}}}
 \def\bnu{{\overline{\nu}}}
 \def\bxi{{\overline{\xi}}}
 \def\bpi{{\overline{\pi}}}
 \def\brho{{\overline{\rho}}}
 \def\bvarrho{{\overline{\varrho}}}
 \def\bsigma{{\overline{\sigma}}}
 \def\bvarsigma{{\overline{\varsigma}}}
 \def\btau{{\overline{\tau}}}
 \def\bphi{{\overline{\phi}}}
 \def\bvarphi{{\overline{\varphi}}}
 \def\bchi{{\overline{\chi}}}
 \def\bpsi{{\overline{\psi}}}
 \def\bomega{{\overline{\omega}}}

\def\thalf{{\textrm{\tiny\textonehalf}}}
\def\tquarter{{\textrm{\tiny\textonequarter}}}
\def\Ko{{\scriptscriptstyle K}}
\def\tKo{\scriptscriptstyle k }
\def\corr{$\clubsuit$}

\newcommand{\auth}{\large J. Greitz\footnote{email:jesper.greitz@kcl.ac.uk}, P.S.\ Howe\footnote{email: paul.howe@kcl.ac.uk}}

\begin{document}

\renewcommand{\thefootnote}{\fnsymbol{footnote}}

\null
\begin{flushright}
{\small KCL-MTH-11-08}\\
\vskip 1.5 cm
\end{flushright}

\begin{center}
{\Large{\bf Maximal supergravity in $D=10$: forms, Borcherds algebras and superspace cohomology}}
\vspace{.75cm}

\auth

\vspace{.5cm}

\begin{center}
{\it\small Department of Mathematics, King's College, London, UK}
\end{center}
\vspace{1cm}

{\bf Abstract}
\end{center}
\vskip .5cm
We give a very simple derivation of the forms of $N=2,D=10$ supergravity  theories  from supersymmetry and $SL(2,\bbR)$ (for IIB). Using superspace cohomology we show that, if the Bianchi identities for the physical fields are satisfied, the (consistent) Bianchi identities for all of the higher-rank forms must be identically satisfied, and that there are no possible gauge-trivial Bianchi identities (i.e.  $dF=0$) except for exact eleven-forms. We also show that the degrees of the forms can be extended beyond the spacetime limit, and that the representations they fall into agree with those predicted from Borcherds algebras. In IIA there are even-rank RR forms, including a non-zero twelve-form, while in IIB there are  non-trivial Bianchi identities for thirteen-forms even though these forms are identically zero in supergravity. It is speculated that these higher-rank forms could be non-zero when higher-order string corrections are included.

\vspace{1cm}



\pagebreak
\tableofcontents
\setcounter{page}{1}


\section{Introduction}


An interesting feature of maximal supergravity theories is that they include $p$-form gauge fields of varying degrees. This set of fields includes some of the physical fields, their duals and further non-physical fields for which the potential forms have degree 
$(D-1)$ and $D$. These sets of forms can be understood in  terms of duality symmetries and supersymmetry alone, but they also have nice algebraic interpretations \cite{Cremmer:1997ct,Cremmer:1998px}. It was subsequently shown that these can be understood in terms of (truncated, super) Borcherds algebras \cite{HenryLabordere:2002dk,HenryLabordere:2002xh}, and also that the spectrum of forms could be obtained  from $E_{11}$ \cite{Julia:1997cy,West:2001as,Riccioni:2007au,Bergshoeff:2007qi,Riccioni:2009xr}. In a recent article \cite{Henneaux:2010ys} it was shown that one can deduce the former from the latter, at least for forms with degrees that do not exceed the spacetime dimension. In this article we shall discuss these related topics in a superspace setting which has the advantage that there is no limit to the degrees of the forms that one can consider. A second advantage is that one can work in a manifestly covariant way in terms of the field strengths and this also allows one to see the algebraic structure in a rather direct fashion, along the lines of that proposed some time ago \cite{Cremmer:1997ct,Cremmer:1998px}.

The superspace approach is straightforward. One starts off with a set of physical forms, including the duals, and then asks how many further forms can be constructed that satisfy consistent Bianchi identities, of the type $dF=F^2$, and that also transform under appropriate representations of the duality group, when present. Here, consistency just means that applying a second $d$ must give zero  on the left of the Bianchi identity and hence also on the right. One thus obtains  an expression cubic in the $F$s on the right which must vanish. Thus, if we couple each field strength to an appropriate generator, the Bianchi identity will determine an antisymmetric product for these generators while the cubic consistency condition gives rise to the Jacobi identity which makes this algebraic structure into a Lie (super)-algebra. 

In general, field strength forms with degree greater than $(D+1)$ vanish in supergravity, but this does not mean that these forms are not of interest. There are some examples of non-vanishing $(D+2)$ forms, including in IIA supergravity in $D=10$, while other forms may have interesting Bianchi identities of the form $dF=F^2$ where the two $F$s on the right-hand side do not vanish even though the left side does identically. An example of this occurs in the IIB theory where there are thirteen-forms whose Bianchi identities involve non-zero lower-degree forms on the right. More importantly, perhaps, it might be that these forms could become non-zero in the presence of higher-order string effects.

In this paper we restrict our attention to maximal supergravity in $D=10$. The full set of forms were constructed explicitly in components in  \cite{Bergshoeff:2006qw} for IIA, in \cite{Bergshoeff:2005ac} for IIB, and in IIB superspace in \cite{Bergshoeff:2007ma}. In \cite{Bergshoeff:2010mv} a minor discrepancy in the earlier component  results was sorted out and the IIA case given in superspace. We begin, in the next section,  by re-examining these forms in a superspace setting and show that the derivations of these results can be considerably simplified by making use of superspace cohomology. In section three we discuss the Bianchi identities beyond the spacetime limit and explicitly exhibit  the $SL(2,\bbR)$ representations (for IIB) and their multiplicities up to degree fifteen. We also include a short discussion of the IIA case up to degree thirteen and show that there is a non-zero RR twelve-form field strength. In section four we briefly review the Borcherds algebras for IIA and IIB and show how the algebra of forms can be understood in terms of this algebraic framework. There are two appendices, on Borcherds algebras and superspace supergravity.

\section{The forms of type II supergravity}


\subsection{IIB}


The bosonic spectrum of the IIB theory consists of the graviton, two scalar fields, the dilaton and axion, a pair of two-form potentials and a four-form potential whose five-form field strength is self-dual. To these we can add their duals, a doublet of seven-form field strengths and a triplet of nine-forms. The latter are dual to the field strengths for the scalars and transform under the triplet representation of $SL(2,\bbR)$ even though there are only two scalars. This can be achieved by means of a constraint on the field strength that ensures that there are only two dynamical dual eight-form potentials. This set can then be extended by a quadruplet and a doublet of eleven-forms, corresponding to the ten-form potentials  studied  in \cite{Bergshoeff:2005ac}.\footnote{ The existence of these forms was known to the authors of  \cite{HenryLabordere:2002dk}; see \cite{julia,paulot}} The set of forms is then $\{F_3^R, F_5, F_7^R, F_9^{RS}, F_{11}^{RST}, F_{11}^R\}$, together with the one-form field strengths for the scalars. The  Bianchi identities for these forms are:

\bea
dF_3^R&=&0\nn\w1
dF_5&=&\ve_{RS} F_3^R F_3^S\nn\w1
dF_7^R&=&F_3^R F_5\nn\w1
dF_9^{RS}&=& F_3^{(R} F_7^{S)}\nn\w1
dF_{11}^{RST}&=& F_3^{(R} F_9^{ST)}\nn\w1
dF_{11}^{R}&=& \ve_{ST} F_3^{S} F_9^{TR} + \frac{3}{4}F_5 F_7^R\ .
\la{1.2}
\eea

The scalar potentials  can be described by an element $\cV_r{}^R$ of $SL(2,\bbR)$ modulo local $U(1)$ gauge transformations, where $r$ is a local $SO(2)$ vector index. The Maurer-Cartan form $d\cV \cV^{-1}=P+Q$, where $Q$ is the $U(1)$ connection and $P$ can be considered  as the one-form field strength for the scalar  potentials. It carries local $SO(2)$ indices and satisfies $DP=0$, but we can convert these indices to global ones by multiplying by two factors of $\cV$ to form the $SL(2,\bbR)$ triplet of one-forms $F_1^{RS}:=\d^{rt}P_t{}^s\cV_r{}^R \cV_s{}^S$. The Bianchi identity for $F_1^{RS}$ is simply $dF_1^{RS}=0$, and indeed one  can solve it by setting  $F_1^{RS}=\half dM^{RS}$, where $M^{RS}:=\d^{rs}\cV_r{}^R \cV_s{}^S$.

It is a simple matter to check that the Bianchi identities \eq{1.2} are indeed consistent.  Furthermore, the full set is determined from the first two (for the physical fields) by consistency and $SL(2,\bbR)$ symmetry.  We shall now show that they can be solved straightforwardly, and that there are no further gauge-trivial identities ($dF=0$). To do this we shall use superspace cohomology which we now briefly explain. 

In superspace the tangent bundle $T$ splits invariantly into even and odd parts, $T_0\oplus T_1$, and it is therefore useful to consider forms with even and odd degrees. Thus the space of $n$-forms, $\O^n$, splits into a sum of spaces of $(p,q)$-forms,  $\O^{p,q}$, where $p+q=n$. In a similar way the exterior derivative splits into components with different bi-degrees:

\be
d=d_0+d_1+ t_0+t_1\ ,
\la{1.3}
\ee

where the bidegrees are $(1,0),(0,1), (-1,2)$ and $(2,-1)$ respectively. The first two, $d_0$ and $d_1$, are essentially even and odd differential operators, while the other two are algebraic operators formed with the dimension-zero and dimension three-halves torsion respectively. In particular,

\be
(t_0 \o_{p,q})_{a_2\ldots a_p \b_1\ldots \b_q}\propto T_{(\b_1\b_2}{}^{a_1}\o_{a_1|a_2\ldots a_p|\b_3\ldots \b_{q+2})}\ ,
\la{1.4}
\ee

where $T_{\a\b}{}^c$ is the dimension-zero torsion which takes its flat space form, i.e. a gamma matrix, in supergravity.

The equation $d^2=0$ splits into various parts according to the bi-degrees amongst which one has

\bea
(t_0)^2&=& 0\la{c4}\w1
t_0 d_1 + d_1 t_0&=&0\la{c5}\w1
d_1^2 +t_0 d_0+ d_0 t_0&=&0\ .
\la{1.5}
\eea

The first of these enables us the define the cohomology groups $H_t^{p,q}$, the space of $t_0$-closed $(p,q)$-forms modulo the exact ones \cite{Bonora:1986ix}. The other two then allow one to define the spinorial cohomology groups $H_s^{p,q}$ \cite{Cederwall:2001bt,Cederwall:2001dx}. These groups make use of a derivative $d_s$ which is essentially $d_1$ acting on $H_t^{p,q}$, but they will not be needed these in this paper. In ten and eleven dimensions these cohomology groups are related to spaces of pure spinors and pure spinor cohomology respectively  \cite{Howe:1991mf,Howe:1991bx,Berkovits:2002zk}.  A key result, which we will make repeated use of below, is that, in $N=2,D=10$ supersymmetry, the groups $H_t^{p,q}$ vanish for $p>1$ \cite{Berkovits:2008qw}. In IIB there are two non-trivial $t_0$-closed $(1,2)$-forms which occur in the solution to the Bianchi identities for the three-forms, whereas in IIA there is only one such form. Using these forms one can construct elements of $H_t^{1,q}$ in terms of $(0,q-2)$ forms.

Now suppose that we have a closed $n$-form whose  lowest-dimensional non-vanishing component (i.e. the one with least even and greatest odd degree)  is $\o_{p,q}$ where $p+q=n$. The first three components of the equation $d\o_n=0$ are

\bea
t_0\o_{p,q}&=&0 \nn\w1
d_1\o_{p,q}+t_0\o_{p+1,q-1}&=&0\nn\w1
d_0\o_{p,q}+d_1\o_{p+1,q-1}+t_0\o_{p-2,q-2}&=&0
\la{1.5.1}
\eea

The lowest component $\o_{p,q}$ is therefore $t_0$-closed, and hence, unless it is exact, will  determine  an element of the cohomology group $H_t^{p,q}$. As we shall see, when combined with the fact that these groups are zero for $p>1$, this makes the analysis of the Bianchi identities rather simple. 

We shall also need some elementary dimensional analysis. In geometrical units (of mass) the dimension of the purely even component of any field strength form $F_n$ is $[F_{n,0}]=1$ (excluding the basis forms) and this implies that $[F_{n-q,q}]=1-\frac{q}{2}$.  In on-shell supergravity there are no scalar or tensor fields with negative dimensions and hence the only non-zero components of any $F_n$ are $F_{n-2,2},\, F_{n-1,1}$ and $F_{n,0}$ with dimensions $0,\frac{1}{2}$ and $1$ respectively. 

Let us write any of the Bianchi identities above in the form

\be
 I^X_{n+1}= dF_{n}^X- (FF)_{n+1}^X\ ,
\la{1.6}
\ee

where $X$ denotes the representation of $SL(2,\bbR)$ under which $F_n$ transforms.  Even if  a particular Bianchi is not satisfied, consistency means that $d I^X_{n+1}=0$  modulo other lower-degree $I$s. Since there are no fields in supergravity that have negative dimensions, the lowest non-vanishing component of any $I$ also has dimension zero and is given by $I^X_{n-3,4}$. This must be $t_0$-closed and will therefore be $t_0$-exact if $n\geq 5$. Now we know that the Bianchi identities are satisfied for all of the physical fields, so we can deduce from this that for all of the other forms, the physical duals and the non-physical eleven forms, the lowest components of the corresponding Bianchi identities are $t_0$-exact.\footnote{This has to be done sequentially so that the lower-dgree $I$s can be ignored.} Thus we will have $I^X_{n-3,4}=t_0 J^X_{n-2,2}$ for some $J^X_{n-2,2}$. But $J^X_{n-2,2}$ has precisely the same index structure as the lowest non-zero component of $F_n$, namely $F^X_{n-2,2}\,$, and hence setting $J^X_{n-2,2}=0$  allows one to solve for  $F^X_{n-2,2}$ in terms of the physical fields  without imposing any further constraints. We can therefore do this and turn our attention to the next level, $I^X_{n-2,3}$, which is also $t_0$-exact and which has the right number of components to allow us to solve for $F^X_{n-1,1}$. Going one step further in a similar fashion we see that we will be able to solve for $F^X_{n,0}$ and hence for the whole of $F_n^X$.

 To make this more explicit we consider the example of the IIB  seven-forms $F_7^R$. We have 

\be
I^R_8= dF^R_7-F_3^R F_5\ ,
\la{1.6.1}
\ee

and we assume that the Bianchi identities for $F_3^R$ and $F_5$ are satisfied. The lowest component of $I_8^R$ that is not trivially satisfied on dimensional grounds is $I^R_{4,4}$. Since $dI_8^R=0$ we have $t_0 I^R_{4,4}=0$, and so, by the vanishing of $H_t^{4,4}$, we find

\be
I^R_{4,4}=t_0 J^R_{5,2}\ .
\la{1.6.2}
\ee

Now

\be
I^R_{4,4}=t_0 F^R_{5,2} - (F^R_3 F_5)_{4,4}\ ,
\la{1.6.3}
\ee

and because of \eq{1.6.2} we know that the second term on the right is itself expressible as $t_0$ acting on some $(5,2)$-form which we can call $[F^R_3\cdot F_5]_{5,2}$. The $(4,4)$ component of this Bianchi identity is clearly solved by  setting $J^R_{5,2}=0$ which means that we can choose $F^R_{5,2}$ to be given by

\be
F^R_{5,2}=[F^R_3\cdot F_5]_{5,2}\ .
\la{1.6.4}
\ee

In principle we could add to this a $t_0$-exact term of the form $t_0 G^R_{6,0}$, but this can be ignored as there are no dimension-zero gauge-invariant possibilities for $G^R_{6,0}$. In other words, the dimension-zero Bianchi identity determines the dimension-zero component of $F^R_7$, i.e. $F^R_{5,2}\,$, in terms of the dimension-zero components of the lower-degree forms which are known quantities. Having solved $I^R_{4,4}$ we can then apply similar arguments to $I^R_{5,3}$ and $I^R_{6,2}$ in order to solve for $F^R_{6,1}$ and $F^R_{7,0}$ in terms of the components of $F^R_3$ and $F_5$.  At the next order, if $t_0 I^R_{7,1}=0$ then automatically $I^R_{7,1}=0$, and similarly for $I^R_{8,0}$. Thus the Bianchi identity $I_8^R$ contains precisely enough information to solve for $F_7^R$ in terms of $F_3^R$ and $F_5$ and does not imply any additional constraints. We can then apply the same analysis to the higher-degree forms in sequence. The upshot of this cohomological analysis is that we do not have to solve for the components of any of the higher-degree forms explicitly in order to verify that solutions to their Bianchi identities are guaranteed to exist. Nevertheless, it is not difficult to find these solutions; they are given explicitly in a local $SO(2)$ basis in \cite{Bergshoeff:2007ma}. The dimension-zero components are constructed from Lorentz- and $SO(2)$-invariant tensors, such as gamma-matrices, the dimension-one-half components are proportional to the physical fermion fields (dilatinos), while the dimension-one components can be physical field-strength tensors or bi-linears in the physical fermions.

To summarise, when the Bianchi identities for the physical fields are satisfied in on-shell IIB supergravity, there is no obstruction to their being solved for the higher-rank form fields provided that the Bianchi identities themselves are formally consistent. In addition, we can show that there are no gauge-trivial forms, i.e. with $dF=0$ Bianchi identities, except with degree eleven. In IIB this could in principle be any of the duals or non-physical forms if it were to turn out that the right-hand sides of any of equations \eq{1.2} could be set to zero. We can see this by applying a similar argument to the one used for the Bianchi identites which implies that the lowest component of such an $F$, $F_{n-2,2}$, has to be exact, $F_{n-2,2}=t_0 G_{n-1,0}$, since $(n-2)>2$. Iterating this one finds that $F$ itself must be exact, $F_n=dG_{n-1}$, say. But the only non-zero component of $G_{n-1}$ is $G_{n-1,0}$ which has to have dimension zero. This can only be some Lorentz-invariant tensor times a function of the scalars, and could therefore only be $\ve_{a_1\ldots a_{10}}$, or $\ve_{10,0}$ in form notation. Thus there is an exact gauge-trivial eleven-form that is a singlet under the duality group. This form is trivial in the sense that it is simply the exterior derivative of the bosonic volume form, and we shall not consider such forms as part of the set of forms that take part in the algebra. Note also that, in particular, this argument  implies that there can be no gauge-trivial eleven-forms in the doublet representation of $SL(2,\bbR)$  in agreement with the superspace discussion in  \cite{Bergshoeff:2007ma}  and the component discussion in  \cite{Bergshoeff:2010mv}.


\subsection{IIA}


The situation in IIA is similar, but there are two differences: there is no duality group and the forms can have both even and odd degree. The physical forms are the RR two- and four-forms, and the NS three-forms; their duals are RR six- and eight-forms and an NS seven-form, together with a nine-form which is dual to the one-form field strength of the dilaton. The RR Bianchi identities, including one for the ten-form, are

\be
dG_{2n+2}=H_3 G_{2n}\qquad {\rm for}\  n=0,1,2,3,4 \ ,
\la{1.7}
\ee

where $G_0$ is taken to be zero  for the standard IIA theory.   The Bianchi identities for the NS forms up to degree nine are

\bea
dH_3&=&0\,, \nonumber \w1
dH_{7}&=& \frac{1}{2} G_4^2 - G_2 G_6\,,\nonumber\w1
dH_9&=&-H_3 H_7 +\frac{1}{2} G_4 G_6 -\frac{3}{2} G_2 G_8\ .
\label{1.8}
\eea

Now consider the possible eleven-form field strengths. There are two
allowable Bianchi identities that can be combined into one:

\bea
dH_{11}&=&A(H_3 H_9 + \frac{3}{2} G_2 G_{10}-\frac{1}{4} G_6^2) \nonumber\\
&\phantom{=}& +B(-G_2 G_{10} + G_4 G_8 -\frac{1}{2}G_6^2)\quad ,
\label{1.9}
\eea

where $A$ and $B$ are real constants.

All of these Bianchi identities are consistent \cite{Bergshoeff:2010mv}, and so we can use the same argument that we used in IIB to prove that they will all be satisfied given that the physical ones are (i.e. those for the two-, three- and four-forms). In IIA $H_t^{p,q}=0$ if $p>1$, and if $p=1$ the basic non-trivial element is $\tilde\C_{1,2}\in H_t^{1,2}$, where $\C_{p,2}$ denotes a symmetric gamma-matrix with $p$ spacetime indices and the tilde indicates the presence of a factor of $\C_{11}$. One possibility for the eleven-forms is that both $A$ and $B$ are zero. In this case we have a gauge-trivial eleven-form, but by the same cohomological argument that we used in IIB, it is exact.

It is easy to verify that there are two gauge  non-trivial eleven-forms by looking at the dimension-zero components. For

\be
H_{9,2}=-iKe^{-2\phi} \C_{9,2}\quad ,
\label{1.10}
\ee

with $K$ constant, we find that \eq{1.9} is satisfied if $2A+8B=K$, so that there are
indeed two independent gauge non-trivial eleven-forms. The other possibility, namely $H_{9,2} \sim i\tilde\C_{9,2}$ corresponds to the gauge-trivial case and so requires $A=B=0$. It is exact, i.e. proportional to $t_0\ve_{10,0}$.

The above analysis can easily be extended to the massive case. To do this one simply has to include a zero-form ``RR'' field strength $G_0=m$ where $m$ is the mass in the Romans deformation of IIA supergravity \cite{Romans:1985tz}. The Bianchi identities for the RR forms take the same form as in \eq{1.7}, but now

\be
 dG_2=G_0 H_3\ ,
\la{1.11}
\ee

while $dH_3=0$ as before. The remaining Bianchi identities also hold provided that one replaces the terms $G_2 G_{2n}\, ,n\geq 3$, with
$G_2 G_{2n}- G_0 G_{2n+2}$. This has been discussed previously in components in \cite{Lavrinenko:1999xi} for forms up to degree nine.


\section{Beyond the spacetime limit}



\subsection{IIB}


We start by considering  the thirteen-forms in IIB supergravity. There are three possibilities corresponding to the five-, three- and one-dimensional representations of $SL(2,\bbR)$:

\bea
dF_{13}^{RSTU}&=& F_3^{(R} F_{11}^{STU)} \nn\w1
dF_{13}^{RS}&=&\ve_{UV} F_3^U F_{11}^{VRS} + \frac{8}{15}F_3^{(R} F_{11}^{S)} + \frac{2}{5}F_5 F_9^{RS} \nn\w1
dF_{13}&=&\ve_{RS} F_3^R F_{11}^S +\frac{3}{8} \ve_{RS} F_7^R F_7^S\ .
\la{2.1}
\eea

Using \eq{1.2} one can easily see that the first of these is consistent. The other two require a bit more work but turn out to be consistent for the given choice of constants. Any thirteen-form must be zero in supergravity since  by dimensional analysis the only possible non-zero components are $F_{11,2}\, ,F_{12,1}$ and $F_{13,0}$ and these all vanish by antisymmetry.   However, all of the forms appearing on the right-hand side of the above equations are non-zero. Moreover, the dimension-zero component of any of the Bianchi identities has the form $I_{10,4}$ and so need not vanish identically.  Nevertheless, since the Bianchi identities are consistent, it follows that  $t_0 I_{10,4}=0$.  This implies that $I_{10,4}=0$ because a cohomologically trivial $(10,4)$-form must vanish  as the superspace has  ten even dimensions. Therefore these Bianchi identities are guaranteed to be satisfied.     The point of this discussion is that it might have been the case that these higher-degree forms should not be considered at all, but these examples give at least a minor indication that it does make sense to include them. One can show by direct computation that the right-hand sides of the $I_{10,4}$ identities vanish, but this  is not trivially obvious unless one invokes the cohomological argument.

Moving on to the fifteen-forms, we find the following set of possibilities

\bea
dF_{15}^{RSTUV}&=& F_3^{(R} F_{13}^{STUV)} \la{2.1.1}\w1
dF_{15}^{RST}&=&a\ve_{UV} F_3^U F_{13}^{VRST} +bF_3^{(R} F_{13}^{ST)} + cF_5 F_{11}^{RST} +  d F_7^{(R} F_9^{ST)}\la{2.1.2}\w1
dF_{15}^R&=&e\ve_{ST} F_3^S F_{13}^{TR}+f F_3^R F_{13} + g F_5 F_{11}^R + h\ve_{ST} F_7^S F_9^{TR}\ .
\la{2.2}
\eea

Applying $d$ to the second of these we find two constraints on the constants $a,b,c,d$ coming from terms with $\ve_{UV} F_3^U F_3^V F_{11}^{RST}$ and $F_5 F_3^{(R} F_9^{ST)}$ so that we can eliminate two of them, say $c$ and $d$. We therefore find that there are two independent fifteen-forms in this representation whose Bianchi identities can be combined into

\bea
dF_{15}^{RST}&=&a(\ve_{UV} F_3^U F_{13}^{VRST} +\frac{5}{8}F_5 F_{11}^{RST} -\frac{5}{8} F_7^{(R} F_9^{ST)})+\nn\w1
&&  + b(F_3^{(R} F_{13}^{ST)} -\half F_5 F_{11}^{RST} +\frac{9}{10} F_7^{(R} F_9^{ST)})\ .
\la{2.3}
\eea

For the doublet representation \eq{2.2} we find three possible consistency conditions from terms with $\ve_{ST} F_3^S F_3^T F_{11}^R$, $\ve_{ST}F_3^S F_5 F_9^{TR}$ and $F_3^R\ve_{ST} F_7^S F_7^T$. However, only two of them are independent and we therefore have two fifteen-forms in the doublet representation. Their Bianchi identities can be written

\bea
dF_{15}^{R}&=&e(\ve_{ST} F_3^S F_{13}^{TR} +\frac{2}{5}F_5 F_{11}^{R})+\nn\w1
&&  + c(F_3^{(R} F_{13} -\half F_5 F_{11}^{R} +\frac{1}{2} \ve_{ST} F_7^{S} F_9^{TR})\ .
\la{2.3}
\eea

The fifteen-forms vanish identically in supergravity, but  not all of the forms on the right-hand side are zero. However, in this case the dimension-zero component of a Bianchi identity has the form $I_{12,4}$ and so must vanish by antisymmetry.

This analysis shows that at each level the number of representations that can arise increases, and that, from degree fifteen onwards, there are also multiplicities in some of the representations.


\subsection{IIA}


For the IIA case we observe first that the Bianchi identities \eq{1.7} for the RR forms are consistent for any value of $n$. For the most part these are trivial in supergravity, but there is a non-zero RR twelve-form with dimension-zero component

\be
G_{10,2}=-iKe^{-\f} \C_{10,2}\ 
\la{2.4}
\ee

for some real non-zero constant K. In fact the Bianchi identity $dG_{12}=H_3 G_{10}$ is automatically soluble, by cohomology, the solution being given by the above expression. (The higher-dimensional components of $G_{12}$ are identically zero.)

One can also have non-trivial thirteen-form Bianchi identities in IIA.  The consistent ones turn out to be

\be
dH_{13}=A(-\frac{5}{4}G_2 G_{12}-\frac{1}{4} G_4G_{10} + \frac{1}{4} G_6G_8) + B(\frac{5}{2} G_2 G_{12}-\frac{3}{2} G_4 G_{10} +\half G_6 G_8)+ H_3 H_{11}\ .
\la{2.5}
\ee

where $A$ and $B$ are the two constants that appear in \eq{1.9}. Since there are two $H_{11}$s depending on the choice of these, there are also two independent thirteen-form Bianchi identities. 

As in the IIB case, the left-hand side of this equation is identically zero in supergravity, and the fact that the right-hand side vanishes as well, even though the individual forms that appear there do not, follows from cohomology.


\section{The Borcherds connection}


The relation of the forms to Borcherds algebras \cite{Borcherds} was discussed in \cite{HenryLabordere:2002dk,HenryLabordere:2002xh}. In ten dimensions, it has been shown that the Borcherds algebra for IIB is the same as one investigated earlier in a different context   \cite{Slansky:1991dx}, while the IIA algebra is a superalgebra. Both algebras have $3\xz 2$ generators. There is also an intriguing relation with del Pezzo surfaces which is discussed in  \cite{HenryLabordere:2002dk,HenryLabordere:2002xh}. In this section we shall relate these algebras to the algebras generated by the forms. The definition of a Borcherds algebra can be found in appendix A.



\subsection{IIB}


The Borcherds algebra for IIB is purely even since all of the field-strengths have odd degree. The Cartan matrix is

\be
\left( \begin{array}{rr}
 0 & -1\\ -1 & 2 
\end{array}\right)\ ,
\la{3.1}
\ee

so that the fundamental commutation relations between the generators are

\bea
[h_0,e_0]&=&0 \qquad  \     \  \  [h_1,e_0]=-e_0 \nn\w1
[h_0,e_1]&=&-e_1\qquad      [h_1,e_1]=2e_1\ .
\la{3.2}
\eea

 There are two more generators $\{f_0,f_1\}$ (see appendix A for the full algebra), and we also have 

\be
(\ad\,e_1)^2 e_0=0\ ,
\la{3.3}
\ee

while $\{f_1,h_1,e_1\}$ forms a basis for $\gs\gl(2)$. The vectors $e_0,e_1$ are eigenvectors associated with the positive simple roots, $\a_0,\a_1$, respectively. 

It is clear from the discussion of the previous section that the algebra of forms is generated from the three-form field strengths. We shall associate a generator with each potential form, so the three-form generators will be denoted $e^2_R$, the five-forms by $e^4$ and so on. In IIB these are all even generators. We write the sum of all the field-strengths as $\bbF=\sum( F_n^X e^{n-1}_X)$ where $X$ denotes the appropriate representation of $\gs\gl(2)$, so that all of the Bianchi identities can be combined into

\be
d\bbF=\half [\bbF, \bbF]\ ,
\la{3.4}
\ee

where the commutator denotes the commutator of the basis elements, and where due care has to be taken with signs.  The generators $e^2_R$ form an $\gs\gl(2)$ doublet, so if we identify the lowest weight $e^2_1$ with $e_0$, the second one can be obtained from it by the raising operator $e_1$, so $e^2_2=[e_0,e_1]:=e_{01}$. This cannot be raised any further so that we have the relation \eq{3.3}, $(\ad\,e_1 )^2e_0=0$. To make further progress we investigate some of the states that are generated. For $F_5$ and $F_7^R$ we have

\bea
e^4&=&[e_0,e_{01}]\nn\w1
e^6_1&=& [e_0,[e_0,e_{01}]]\nn\w1
e^6_2&=& [e_{01},[e_0,e_{01}]]\ .
\la{3.5}
\eea

Continuing in this way we find, for form degree $2n+1$, a state of the form $(\ad\, e_0)^{n-1} e_{01}$, and this series can increase without limit. Moreover, it is clear that each state will be characterised by a corresponding root, although one should be aware that these can occur with multiplicities. Each $(2n+1)$-form is associated with the roots $\a=n\a_0 +m\a_1$ where $m=1,\ldots (n-1)$, although there can be multiplicities starting from $n=5$. It is easy to see that one recovers the previously known results up to level 5, i.e. the forms of degree eleven that saturate the spacetime limit (ten-form potentials). However, the positive roots have been tabulated beyond this level \cite{Slansky:1991dx}, so that we can easily compare our results from section three to the Borcherds prediction up to level 7, i.e. fifteen-forms. The table is as follows:

\begin{center}
\begin{tabular*}{0.75\textwidth}{@{\extracolsep{\fill}} |c|c|l|}
\hline
Level & Form degree& $\gs\gl(2)$ representation(s) \\
\hline\hline
1 & 3& 2\\
\hline
2&5&1\\
\hline
3&7&2\\
\hline
4&9&3\\
\hline
5&11&4+2\\
\hline
6&13&5+3+1\\
\hline
7&15&6+4(2)+2(2)\\
\hline
\end{tabular*}
\end{center}

\vskip .5cm

The form degrees here are those of the field strengths while the figures in brackets in the last entry in the third column indicate that these representations appear with multiplicity two. Comparing with the Bianchi identities in section three we find exact agreement including the correct multiplicities for the fifteen-forms. It is also easy to see that the roots  are correctly given. For example, the vector $(\ad\,e_0)^n e_1$ is the lowest weight state of the largest representation at level $n$ corresponding to the root vector $n\a_0 + \a_1$. 
Indeed, this result is not surprising because it is clear that the algebra of forms must be isomorphic to the positive root algebra $\cN^+$ modulo the one-dimensional space generated by $e_1$.

It is also clear that the Borcherds algebra determined by \eq{3.2} is the smallest Borcherds algebra that can accommodate the IIB form algebra. The existence of an $sl(2)$ subalgebra implies that $a_{11}=2$ while \eq{3.3} tells us that $a_{01}=-1$ (and hence, by symmetry, that $a_{10}=-1$). The fact that one can have arbitrary powers of $ad\, e_0$ means that $a_{00}$ cannot be positive. If it was negative there would be a second $sl(2)$ subalgebra with infinite-dimensional representations within the Borcherds algebra, but this is not possible because there is only a finite number of forms of a given degree. So $a_{00}=0$ and we are thus led to the Cartan matrix \eq{3.2}.


\subsection{IIA}


The Cartan matrix for IIA is given by

\be
\left( \begin{array}{rr}
 0 & -1\\ -1 & 0
\end{array}\right)\ .
\la{3.6}
\ee

The super-algebra has generators $\{f_0,f_1,h_0,h_1,e_0,e_1\}$, where $e_1,f_1$ are odd, which obey the basic commutation relations

\bea
[h_0,e_0]&=&0\nn\w1
[h_0,e_1]&=&-1\nn\w1
[h_1,e_0]&=&-1\nn\w1
[h_1,e_1]&=&0\ .
\la{3.7}
\eea

Since $a_{00}=a_{11}=0$, the subalgebras associated with both sets of generators  are of Heisenberg type. 

The form algebra is generated from $G_2$ and $H_3$ so we shall associate elements of this algebra with them. For $G_2$ this is the odd element, $e_1$, while for $H_3$ it is the even element $e_0$.  The first point to notice is that the $e_0$ component of \eq{3.4} is

\be
dH_3\, e_0=\half G_2 G_2 [e_1,e_1]\ .
\la{3.8}
\ee

For this to agree with the correct identity, i.e.  $dH_3=0$, we must have the relation $[e_1,e_1]=0$. But this is also required from the general rules for a Borcherds superalgebra in the appendix. 

For the RR forms the situation is very simple. $G_4$ is associated with $[e_0,e_1]:=e_{01}$, $G_6$ with $[e_0,e_{01}]:=e_{001}$ and so on. For $G_{2n}$ the element of the algebra is $(\ad\, e_0)^{n-1} e_1$, and this series can increase without limit since $a_{00}=0$.

For the NS forms the situation is slightly more complicated, but can be obtained directly from the Bianchi identities. For the seven-form one has only one possibility, namely $[e_{01},e_{01}]$, while for the nine-form one has $[e_0,[e_{01},e_{01}]]$. However, for the eleven-forms one finds two possibilities, $[e_0,[e_0,[e_{01},e_{01}]]]$ and $[e_{001},e_{001}]$. (Note that all of the vectors $e_{00..1}$ for any number of zeroes are odd.) For the thirteen-forms there are again two possibilities, $(\ad\,e_0)^3 [e_{01},e_{01}]$ and $\ad\,e_0[e_{001},e_{001}]$. These results are in agreement with those of section three. This pattern continues to higher levels, so that there is a series of terms of this type obtained by acting with $\ad\, e_0$ on vectors of the form $[(\ad\,e_0)^k e_1,(\ad\,e_0)^k e_1]$. The situation can be summarised rather simply. For the RR forms, $G_{2n}$, $n\geq1$, the roots are $(n-1)\a_0 + \a_1$, all with multiplicity one. For the NS forms, one has $H_3$ with root $\a_0$ and two series (both with $n\geq1$): the $(4n+3)$-forms, which correspond to the roots $2n\a_0+2\a_1$, and which have multiplicity $n$, and the $(4n+5)$-forms, which correspond to the roots $(2n+1)\a_0 + 2\a_1$ and which also have multiplicity $n$.  These results for the multiplicities can be seen directly by inspection or by means of  the Peterson formula for super-Borcherds algebras which is briefly summarised in appendix A. 

In the IIA case the algebra of forms is clearly isomorphic to the positive part, $\cN^+$, of the Borcherds superalgebra. It is also clear that this Borcherds superalgebra is the smallest one that can accommodate the form algebra. Since $[e_1,e_1]=0$ it follows that $a_{11}=0$ while $a_{00}$ must be zero for similar reasons to the IIB case.  It is then possible to normalise the generators so that $a_{01}=a_{10}=-1$.


\section{Discussion}


In this article we have seen that the use of superspace techniques for maximal supergravity in $D=10$ simplifies the discussion of the forms in the theory. Because the odd basis forms in superspace are commutative there is no limit to the degrees that forms can have. This means that we can avoid the use of potentials for the ten-forms, and that we can continue beyond the spacetime limit. The resulting formalism is manifestly covariant and automatically determines a Lie-(super)algebraic structure because the forms satisfy Bianchi identities of the form $dF=F^2$, which determines a graded antisymmetric multiplication, while the consistency conditions for the Bianchis are equivalent to the Jacobi identity. The algebras determined by the infinite set of possible forms are infinite-dimensional and are given by the positive elements of Borcherds algebras.  In the IIB case the generator $e_1$ is associated to the axion  \cite{HenryLabordere:2002dk,HenryLabordere:2002xh}. 

In  \cite{Henneaux:2010ys} it was shown that these algebras, for all maximal supergravities, can be derived from $E_{11}$ after suitable tensoring with an appropriate Grassmann algebra. However, it is not so clear how the higher-degree forms can be accommodated in this formalism.  It might be that there is some extension of $E_{11}$ that could account for them, or it might be the case that these forms should  simply be disregarded.

In $N=2, D=10$ supergravity nearly all of the forms with degree greater than eleven are zero, the exception being the IIA RR twelve-form that was mentioned in section three. However, it might be that these forms could be non-zero when string-theory corrections are switched on. This is not an easy problem to investigate because one is faced  with group-theoretic difficulties in whichever dimension one chooses to work with. In $D=10$, for example, one could have $\a'^3$ corrections in the $(0,13)$ components of the thirteen-forms that would have to be linear in the dilatinos. The problem here is that one is faced with representations of the spin group involving the tensor product of thirteen spinors. On the other hand, in $D=3$ there can be non-zero five-forms \cite{Greitz:2011vh} whose lowest components would have to be  $\a'^3$ multiplied by dimension nine-halves functions of the fields, again not easy to analyse.

The fact that one can have forms with higher degrees than the spacetime dimension is implicit in the construction of the hierarchies of forms that appear in the context of gauged maximal supergravity theories in lower-dimensional spacetimes. Indeed, the possibility that the hierachy could in principle be continued indefinitely was mentioned in \cite{deWit:2008gc}. However, it is only in the superspace context that the Bianchi identities for these forms really make sense, and only in superspace that such forms can actually be non-zero.  Perhaps the simplest example of this occurs in the maximal $D=3$ gauged theory where there are five-forms that are non-zero even in the lowest-order theory \cite{Greitz:2011vh} and which are necessary in the gauged Bianchi identites. 

The higher-degree forms would not seem to have any sort of brane interpretation. For example, the notion of an eleven-brane in ten-dimensional spacetime seems rather counterintuitive. On the other hand, the fact that there is a non-zero RR twelve-form might lead one to ask whether it could have any geometrical significance beyond its algebraic r\^ole in the Borcherds context.

\pagebreak

\vskip .5cm

{\bf\Large{Acknowledgements}}

\vskip .5cm

We thank Teake Nutma for helpful comments.

JG thanks the  Marcus Wallenberg Foundation and the STFC for financial support.

\vskip .5cm

 \appendix
 
 {\bf \Large{Appendices}}


\section{Borcherds algebras}


The definition of a Borcherds (or generalised Kac-Moody) (super)-algebra starts with a generalised symmetric Cartan matrix, $(a_{ij}),\ i.j=1\dots N$, where some subset of the indices can be odd, which is non-degenerate and for which the following rules hold. The diagonal elements $a_{ii}$ (no sum) can be positive, negative or zero, while the off-diagonal elements, $a_{ij},\ i\neq j$, are less or equal to zero. In the case that $a_{ii}>0$, then $\frac{2a_{ij}}{a_{ii}}\in\bbZ, \forall j$, while if $i$ is also odd $\frac{a_{ij}}{a_{ii}}\in\bbZ, \forall j$.

The Borcherds algebra $\cA$ associated with $(a_{ij})$ is then determined by $3N$ generators $\{h_i,e_i,f_i\}$, $i=1\ldots N$, satisfying the following conditions:

\bea
[h_i,h_j]&=&0   \la{A.1}\w1
[h_i,e_j]&=&a_{ij} e_j,\qquad [h_i,f_j]=-a_{ij} e_j,\qquad [e_i,f_j]=\d_{ij} h_i \la{A.2}\w1
(\ad\, e_i)^{1-\frac{2a_{ij}}{a_{ii}}}e_j&=&0,\qquad {\rm for}\  \ a_{ii}>0\ {\rm and}\  i\neq j\la{A.3}\w1
[e_i,e_j]&=&0\qquad {\rm when} \ \ a_{ij}=0\la{A.4}\ ,
\eea

with the last two conditions remaining valid if $e_i,\,e_j$ are replaced by $f_i,\,f_j$. The generators $h_i$ are even, and the generator $f_i$ is even or odd if $e_i$ is. If $a_{ii}>0$ the integer $\frac{2a_{ij}}{a_{ii}}$ is negative, and if $i$ is odd, it is also even. 

In a Borcherds algebra there is still a triangular decomposition of the form $\cA=\cN^-\oplus\cH\oplus \cN^+$, and it is still possible to define roots as in the Kac-Moody case. Furthermore, if $a_{ii}>0$, the algebra generated by $\{f_i,h_i,e_i\}$ for $i$ even, or by these together with $[f_i,f_i]$ and $[e_i,e_i]$ when $i$ is odd, are isomorphic to $\gs\gl(2)$ or $\go\gs\gp(1|2)$, respectively, and the algebra can be decomposed into finite dimensional representations of these (super)algebras. When $a_{ii}<0$, one has the same algebras but the Borcherds algebra contains infinite-dimensional representations of them. In the case that $a_{ii}=0$, the sub-algebra generated by $\{f_i,h_i,e_i\}$ is isomorphic to the Heisenberg (super)algebra.

The multiplicities of the roots for Borcherds algebras may be computed using the Peterson formula \cite{peterson,kac}. We found the discussion in \cite{kangkwonoh} to be useful, especially for the super case. Let $\b$ be an element of the positive root lattice $Q^+$, i.e. a linear combination of the positive simple roots (in both of the cases we discuss in the text,  $\a_0$ and $\a_1$,) with non-negative integral coefficients. And let $\gg_{\b}$ be the subalgebra of the Borcherds algebra corresponding to a root $\b$. The super-dimension of such a subalgebra is defined by ${\rm sdim}\, \gg_{\b}=(-1)^{{\rm deg}\,\b} {\rm dim}\, \gg_{\b}$, where the degree is zero or one according to whether $\b$ is even or odd. The ordinary dimension is the multiplicity.  The Peterson formula is

\be
(\b|\b-2\r) c(\b)=\sum (\b'|\b'') c(\b') c(\b'')\ ,
\la{a.2}
\ee

where the sum is over all elements such that $\b=\b' + \b''$, $\r$ is a special combination of the simple roots and the quantities $c(\b)$ are determined by the formula

\be
c(\b)=\sum_n \frac{1}{n} {\rm sdim}(\frac{\b}{n})\ ,
\la{a.3}
\ee

$n$ being a positive integer. The quantity $\r$ is determined by requiring that the left-hand-side of \eq{a.2} should be zero for the positive simple roots; for IIB, $\r=-\a_0$ while for IIA, $\r=0$. The round brackets denote the scalar product determined by the Cartan matrix, with $(\a_i|\a_j)=a_{ij}$ for the positive simple roots.  In the sum in \eq{a.3} the dimension of an element $\b$ of $Q^+$ that is not a root is zero, although this does not mean that the corresponding $c(\b)$ vanishes because $\b$ may be a multiple of a root.  Note that \eq{a.3} only has more than one term if $\b$ is an integral multiple of a root. The $c(\b)$s, and hence the multiplicities, can be computed from these two formulae in an iterative fashion.


\section{Superspace supergravity}


Although the details of the superspace descriptions of $N=2, D=10$ supergravities are not needed for the discussion in the main text we collect here a few basics and references.

The complete IIB supergravity (for the physical fields) was written down in superspace in \cite{Howe:1983sr}, the component version having been given in   \cite{Schwarz:1983qr}. The dual forms were added in \cite{Cederwall:1996ri,Dall'Agata:1998va}, and all of the forms up to degree eleven in \cite{Bergshoeff:2007ma}. The conventions we follow here are those of \cite{Bergshoeff:2007ma} although we have slightly changed the normalisations of some of the forms and written them with upper $SL(2,\bbR)$ indices. One can transform from these to those of \cite{Bergshoeff:2007ma} by means of the $\ve$-tensor. In the original paper a complex $U(1)$ notation was used for the spinors, but it is probably more convenient to use a real $SO(2)$ notation, as in \cite{Berkovits:2001ue}, where the relation between the conventions of \cite{Howe:1983sr} and \cite{Bergshoeff:2007ma} can be found. 

The dimension-zero torsion is given by

\be
T_{\a i\b j}{}^c=i\d_{ij}(\c^c)_{\a\b}\ ,
\la{b1}
\ee

where $i,j=1,2$ are $SO(2)$ spinor indices (we use $r,s,\ldots$ for vector indices). The geometric tensors cannot contain the scalar fields, as the formalism is $SL(2,\bbR)$ covariant, as well as having a local $U(1)$ symmetry for which the gauge field is a composite constructed from the scalars in the usual manner. The spin one-half fields are found in the dimension-one-half torsion, while the other physical field strengths arise at higher dimension, although the (bosonic) five-form does not appear directly. 

In the text we gave the forms in an $SL(2,\bbR)$ basis, but it is sometimes convenient to use the $SO(2)$ basis, the two being related by the scalar matrix $\cV_r{}^R$. In this basis the Bianchi identities take the form

\be
D F=FF - F\wedge P
\la{b2}
\ee

where $D$ is covariant with respect to $SO(2)$, $FF$ denotes the same term that occurs in the $SL(2,\bbR)$ basis, except that the indices are now lower case, and $P$ denotes the matrix of one-forms in the  representation appropriate to the form $F$ on the left.  One advantage of this basis is that the scalars cannot appear undifferentiated so that the dimension-zero components are simply given by products of ($16\xz 16$) gamma-matrices and $SO(2)$ gamma-matrices, $(\t^r)_{ij}$ (symmetric, traceless), $\d_{ij}$ or $\ve_{ij}$.  We refer the reader to \cite{Bergshoeff:2007ma} for details. 

The IIA theory was written down in components in \cite{Campbell:1984zc} and in superspace in \cite{Carr:1986tk}. It was also derived by superspace dimensional reduction from  $D=11$ in \cite{Howe:2004ib}.\footnote{In appendix C of  \cite{Howe:2004ib} equation (C.7) should read $\chi_\a=i(\C_{11}\nab)_\a\F$.}  The version we use here was briefly outlined in \cite{Bergshoeff:2010mv}. We use thirty-two component Majorana spinors. The dimension-zero torsion is

\be
T_{\a\b}{}^c=-i(\C^c)_{\a\b}\ .
\la{b3}
\ee

The string frame is used, so that the dimension-zero component of $H_3$ has no factor of the dilaton, $H_{1,2}\propto \tilde\C_{1,2}$, while the dimension-zero components of the RR forms all have a factor of $e^{-\f}$, multiplied by appropriate gamma-matrices. Since $dG_{2n+2}=H_3 G_{2n}$, this implies that the dimension-zero components of the RR forms have a factor of $\C_{11}$ for $n$ even, but not for $n$ odd.


\end{document}